\documentclass[sigconf]{acmart}

\AtBeginDocument{%
  \providecommand\BibTeX{{%
    \normalfont B\kern-0.5em{\scshape i\kern-0.25em b}\kern-0.8em\TeX}}}

\setcopyright{acmlicensed}
\copyrightyear{2026}
\acmYear{2026}
\acmDOI{XXXXXXX.XXXXXXX}

\acmConference[2026]{Make sure to enter the correct
  conference title from your rights confirmation email}
\acmISBN{978-1-4503-XXXX-X/18/06}

\begin{document}

\title{When Phase Doesn\textquoteright{}t Matter: Self-Coherent Over-the-Air Computation at Sub-THz}

\author{Sherif Ghozzy}
\affiliation{%
  \institution{Department of Electrical and Computer Engineering, Princeton University}
  \country{USA}}
\author{Mohamed Seif}
\affiliation{%
  \institution{Department of Electrical and Computer Engineering, Princeton University}
  \country{USA}}
\author{H.~Vincent Poor}
\affiliation{%
  \institution{Department of Electrical and Computer Engineering, Princeton University}
  \country{USA}}
\author{Kaushik Sengupta}
\affiliation{%
  \institution{Department of Electrical and Computer Engineering, Princeton University}
  \country{USA}}


\renewcommand{\shortauthors}{Sherif Ghozzy, Mohamed Seif, H. Vincent Poor, and Kaushik Sengupta}

\begin{abstract}
Over-the-air computation (OAC) enables efficient function aggregation in wireless networks by
exploiting the superposition property of the multiple-access channel. However, practical
deployment of OAC is severely challenged by the reliance on accurate carrier synchronization
and coherent reception, which are costly and fragile, especially in short-range and
low-complexity systems. In this work, we propose a \emph{self-coherent, synthesizer-free
over-the-air computation framework} based on \emph{Kramers--Kronig (KK) reception}. By
transmitting a biased aggregate waveform and employing direct detection followed by KK phase
reconstruction at the receiver, the proposed scheme eliminates the need for explicit carrier
recovery while preserving coherent-like signal aggregation.

We develop a signal-domain system model for multi-user OAC under KK reception and provide a
synchronization-relaxation analysis demonstrating that the proposed architecture fundamentally
removes carrier-frequency offset (CFO) sensitivity between transmitters and receiver. By
shifting synchronization complexity away from strict carrier-phase tracking and eliminating
distributed phase alignment requirements, the framework reduces control overhead and improves
scalability in multi-user aggregation. A detailed per-symbol mean-squared error (MSE)
characterization isolates the impact of channel mismatch and KK reconstruction noise, showing
that the proposed self-coherent architecture approaches the theoretical performance limits of
baseband OAC under practical operating conditions. Finally, we demonstrate that the approach
is particularly well suited for mmWave and sub-THz systems, where oscillator phase instability
otherwise represents a fundamental bottleneck to scalable coherent OAC.
\end{abstract}




\maketitle

\section{Introduction}
\label{sec:intro}

Scaling millimeter-wave (mmWave) and sub-terahertz wireless systems is critical for latency-sensitive, data-intensive applications, yet remains fundamentally constrained by synchronization complexity. While higher carrier frequencies offer abundant bandwidth and enable highly directional links, they exacerbate carrier frequency offsets (CFO), phase noise, and hardware non-idealities. Conventional spectrally efficient modulation relies on fully coherent receivers, requiring tight frequency and phase alignment between transmitters and receivers. Maintaining such coherence incurs significant overhead, latency, and system complexity, posing a major obstacle to scalable low-latency wireless networks.

Next-generation wireless systems are increasingly expected to support distributed intelligence at scale. Applications such as federated learning and distributed sensing require frequent aggregation of high-dimensional updates from many spatially distributed nodes \cite{seif2025collaborative}. In conventional uplink architectures, aggregation is performed digitally via orthogonal multiple access followed by decoding at a central server. While robust, this separation scales poorly, increasing latency, reducing spectral efficiency, and amplifying synchronization overhead.

Over-the-air computation (OAC) addresses this limitation by leveraging wireless superposition to compute functions directly in the analog domain \cite{csahin2023survey}. When multiple users transmit simultaneously, the channel naturally performs a weighted summation, enabling direct analog aggregation with latency independent of the number of nodes. As a result, OAC has emerged as a promising primitive for communication-efficient distributed learning \cite{you2023broadband,pradhan2025experimental}.

In coherent OAC, transmitters pre-equalize channels such that their superposition yields the desired function. However, this relies on accurate channel inversion and tight carrier-phase alignment across all users \cite{shao2021federated}. Residual (CFO), phase noise, imperfect channel inversion, and timing misalignment introduce phase drift that accumulates across users and symbols. As shown in prior experimental and theoretical studies, even small phase errors can severely degrade aggregation fidelity, particularly in wideband OFDM systems and at high carrier frequencies \cite{csahin2025feasibility,shao2021federated}. Existing approaches attempt to mitigate these impairments through synchronization protocols, phase tracking, or digital compensation \cite{you2023broadband,shao2021federated}, but at the cost of increased overhead or receiver complexity.

Existing digital OAC schemes further exhibit trade-offs in bandwidth efficiency, power consumption, synchronization requirements, and CFO sensitivity, as summarized in Table~\ref{tab:oac_comparison}. Most approaches either require strict phase alignment or sacrifice spectral efficiency and robustness, highlighting synchronization as the central bottleneck in scalable OAC.

In this work, we argue that this limitation is architectural. Rather than enforcing increasingly stringent carrier-phase alignment across distributed transmitters, we eliminate the need for coherent phase tracking at the receiver. Specifically, we introduce a self-coherent Kramers--Kronig (KK) mixed-signal reception framework as a synchronization enabling primitive for multi-user over-the-air computation.

The key idea is to shift the aggregation paradigm from phase-coherent summation in the complex field to envelope-based reconstruction under a minimum-phase constraint. By leveraging envelope detection followed by logarithmic processing and Hilbert-transform-based phase reconstruction, the receiver recovers the complex field deterministically without requiring carrier-phase synchronization with each transmitter. In this architecture, residual CFO and phase offsets do not manifest as destructive multi-user interference, but are instead absorbed into the reconstruction through the structure of the received signal.

This perspective is particularly well suited for mmWave and sub-THz systems, where links are highly directional and bandwidth-rich, but synchronization across distributed oscillators is fragile and costly. By shifting the synchronization burden away from tight carrier-phase locking and toward lightweight transmitter-side alignment, the proposed approach enables scalable aggregation without proportional overhead or receiver complexity.

\begin{table}[t]
\centering
\caption{Comparison of Digital OAC Schemes}
\label{tab:oac_comparison}
\renewcommand{\arraystretch}{1.1}
\setlength{\tabcolsep}{4pt}
\resizebox{0.5\textwidth}{!}{%
\begin{tabular}{p{3.2cm} p{2.6cm} p{2.2cm} p{3.0cm} p{1.8cm}}
\toprule
\textbf{Scheme} 
& \textbf{BW Efficiency} 
& \textbf{Power} 
& \textbf{Synchronization} 
& \textbf{CFO} \\
\midrule

One-bit \cite{zhu2020one} 
& High 
& Moderate 
& Strict 
& High \\

FSK \cite{csahin2021distributed}
& Low 
& Low 
& Relaxed 
& Moderate \\

OFDM 
& High 
& High 
& Strict 
& High \\

\textbf{This Work} 
& High 
& Low
& No carrier recovery
& Low \\

\bottomrule
\end{tabular}}
\end{table}

\subsection{Related Work}

Prior work has examined the impact of phase impairments on coherent OAC and federated learning systems. In~\cite{dahl2024over,dahl2024over2}, oscillator phase noise is analytically characterized, showing how Wiener phase drift within a coherence block leads to non-uniform degradation across transmitted gradient symbols. To mitigate this effect, gradient permutation strategies such as roll, flip, and importance-based sorting are proposed to prioritize more critical updates earlier in transmission. Complementary work in~\cite{dahl2024over2} studies reciprocity-calibrated coherent OAC under phase misalignment, proposing detection and selective realignment policies to balance aggregation accuracy and synchronization overhead. These results highlight that coherent superposition requires tight carrier-phase alignment and becomes increasingly sensitive to phase drift at higher carrier frequencies.

In contrast, the self-coherent receiver architecture proposed in this work fundamentally relaxes the need for transmitter--receiver carrier-phase alignment. By reconstructing the complex field from the received envelope under a minimum-phase constraint, the receiver inherently resolves CFO between distributed transmitters and the receiver, irrespective of RF carrier frequency. As a result, aggregation fidelity is no longer limited by inter-device carrier-phase coherence, even at mmWave and sub-THz frequencies.

Local oscillator (LO) phase noise remains a practical hardware impairment. While the proposed architecture removes the need for strict distributed synchronization, prior techniques such as gradient scheduling, permutation strategies, and selective phase realignment~\cite{dahl2024over,dahl2024over2} can be incorporated to further enhance robustness. These approaches are complementary, addressing residual LO instability, while the proposed architecture eliminates the architectural dependence on coherent carrier alignment.

These challenges are particularly pronounced in multiple-access scenarios, where many distributed devices transmit simultaneously over a shared wireless medium. OAC leverages wireless superposition for fast aggregation, but its performance critically depends on coherent transmission and reception across nodes \cite{csahin2023survey}. This requires accurate carrier synchronization, phase tracking, and channel estimation both between transmitters and the receiver and implicitly across transmitters. At mmWave and sub-THz frequencies, these requirements become increasingly difficult to satisfy, limiting the practicality of OAC.

Prior work has explored algorithmic and receiver-side techniques to relax synchronization requirements, including matched filtering, oversampling, and digital compensation~\cite{shao2021federated}. However, these approaches typically assume coherent downconversion at the receiver, which remains a major bottleneck at high carrier frequencies due to the need for power-hungry synthesizers, tight phase-locked loops, and long acquisition times. Consequently, synchronization overhead and receiver complexity scale poorly with carrier frequency and network size.

In this work, we propose self-coherent Kramers--Kronig (KK) mixed-signal processing as a fundamentally different approach to synchronization in multiple-access systems. By broadcasting a reference carrier alongside the information-bearing signal, self-coherence is established without requiring a high-frequency local oscillator or explicit carrier recovery at the receiver. Envelope detection followed by KK-based processing enables phase recovery from amplitude measurements, achieving coherent-like demodulation while eliminating the most challenging transmitter--receiver synchronization requirements. The remaining synchronization burden is confined to alignment among transmitters, which is significantly easier to manage.

The underlying principle of extracting phase from amplitude traces follows from the Kramers--Kronig relations, which enforce a Hilbert-transform relationship under appropriate signal conditions~\cite{voelcker2003demodulation}. While prior KK receivers in optical systems rely on high-speed ADCs and power-intensive digital processing~\cite{harter2020generalized}, our approach employs low-power analog and mixed-signal implementations~\cite{ghozzy202412}, enabling energy-efficient operation for wireless systems.

\subsection{Contributions}

The main contributions of this work are as follows:

\begin{itemize}
\item \textbf{Self-coherent aggregation architecture:} A mixed-signal Kramers--Kronig reception framework enabling multi-user OAC without strict carrier-phase synchronization.

\item \textbf{Synchronization-relaxation analysis:} System-level analysis showing that self-coherent KK reception fundamentally relaxes CFO and phase alignment constraints in conventional coherent OAC.

\item \textbf{Low-overhead multi-user scalability:} Reduced control overhead and latency by eliminating distributed phase tracking and shifting synchronization complexity away from the receiver.

\item \textbf{Relevance to high-frequency networks:} Alignment with mmWave and sub-THz operation, where phase instability is the dominant scalability constraint.

\end{itemize}

Overall, this work positions self-coherent KK mixed-signal processing as a systems-level primitive enabling scalable over-the-air computation in next-generation distributed wireless networks.



\section{Problem Statement \& System Model}
\label{sec:system_model}
This section establishes the system model used to analyze the proposed
self-coherent KK receiver in the context of OAC. We begin by contrasting conventional coherent receivers with
the proposed self-coherent architecture, highlighting the synchronization
constraints that limit scalable OAC at high carrier frequencies. We then examine
the class of nomographic functions supported by the KK-based framework and
evaluate the achievable aggregation performance through system-level
simulations. Finally, we introduce the transmission scheme and signal model used
throughout the paper, including the waveform construction, channel model, and
the discrete-time equivalent representation of the KK reconstruction process.

\begin{figure*}[!ht]
  \centering
  \includegraphics[width=\linewidth]{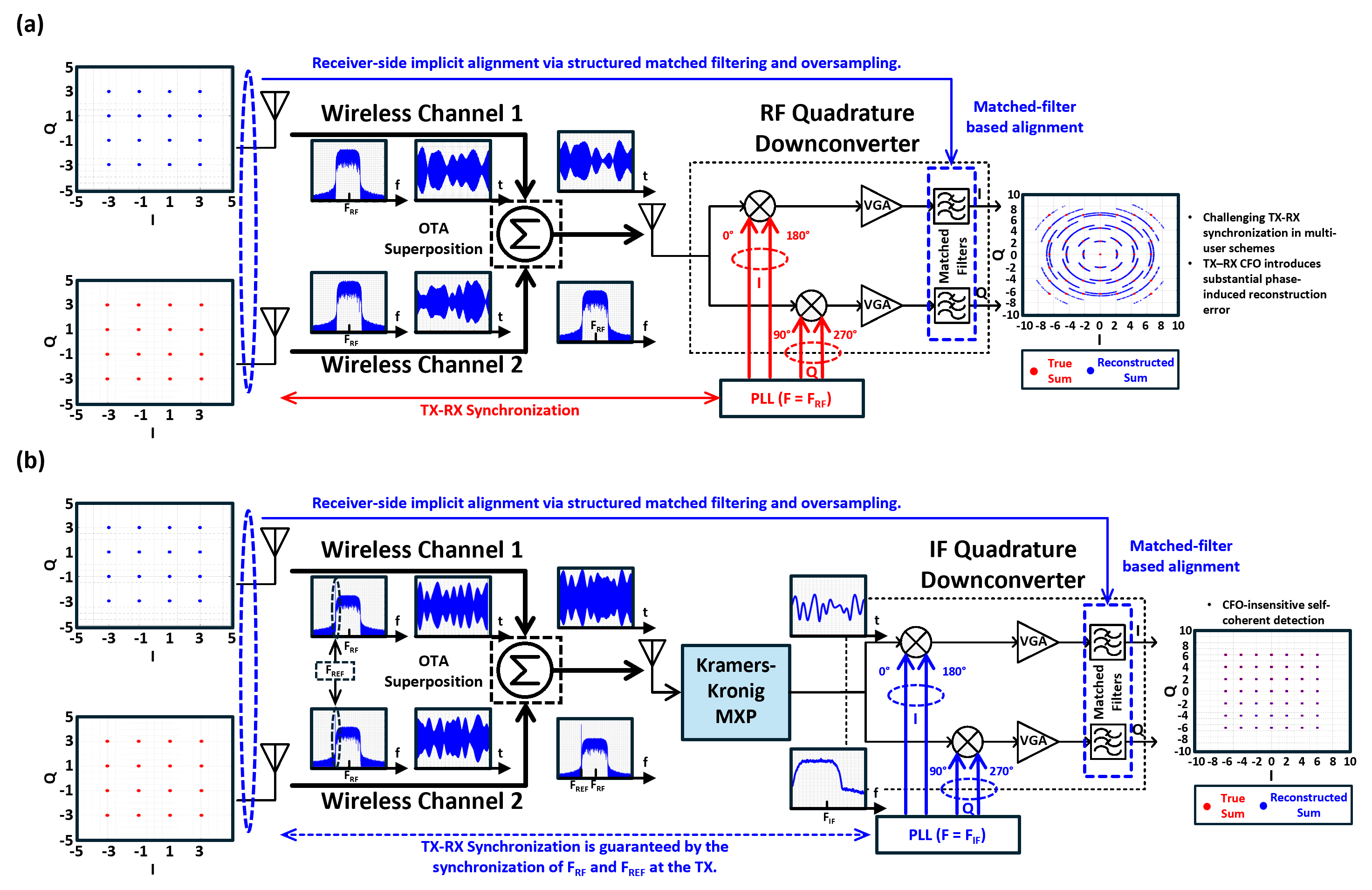}
  \caption {Over-the-air computation system model for two users transmitting concurrently around the carrier frequency $F_{\mathrm{RF}}$. (a) Conventional direct-conversion reception after RF superposition of the users\textquoteright{} signals, where accurate synchronization is required both among the transmitted signals and between the transmitter and receiver local oscillators to correctly recover the computed sum. (b) Self-coherent Kramers--Kronig reception, where synchronization is only required among the transmitted signals, as the receiver is inherently self-coherent through the broadcasted reference carrier.}
  \label{SystemModel}
\end{figure*}
\begin{table}[!t]
\centering
\caption{System Parameters}
\label{tab:sim_params}
\resizebox{0.5\textwidth}{!}{%
\begin{tabular}{l l l}
\hline
\textbf{Category} & \textbf{Parameter} & \textbf{Value} \\
\hline
Data Source & Binary distribution & Bernoulli, $P\{0\}=0.5$ \\
& Initial seed & Fixed (seed = 2) \\

Modulation & Modulation scheme & Rectangular $M$-QAM \\
& Bits per symbol & $N_{\mathrm{bps}} = 4$ \\
& Constellation size & $M = 2^{N_{\mathrm{bps}}}$ \\
& Mapping & Gray coding \\
& Phase offset & $0$ rad \\

Rates & Symbol rate & $R_s = 1$ GHz \\
& Bit rate & $R_b = N_{\mathrm{bps}} R_s$ \\

Sampling & Baseband sampling rate & $f_s = n_{\mathrm{tsteps}} / T_s$ \\
& Samples per symbol & $rrc_{\mathrm{sps}}$ \\

Pulse Shaping (Tx) & Filter type & Square-root raised cosine \\
& Roll-off factor & $\alpha = 0.35$ \\
& Filter span & $10 \times rrc_{\mathrm{sps}}$ symbols \\
& Gain & $\sqrt{rrc_{\mathrm{sps}}}$ \\

Pulse Shaping (Rx) & Filter type & Square-root raised cosine (matched) \\
& Decimation factor & $rrc_{\mathrm{sps}}$ \\
& Decimation offset & DO = 1 \\
& Gain & $1/\sqrt{rrc_{\mathrm{sps}}}$ \\

Power / Noise & SNR & 30 dB \\
& CSPR & 6 dB \\

Simulation & Number of symbols & $N_{\mathrm{symb}} = 100000$ \\
& Simulation duration & $T_{\mathrm{sim}} = N_{\mathrm{symb}} N_{\mathrm{bps}} T_b$ \\
\hline
\end{tabular}}
\end{table}

\begin{figure}[!t]
  \centering
  \includegraphics[width=1\linewidth]{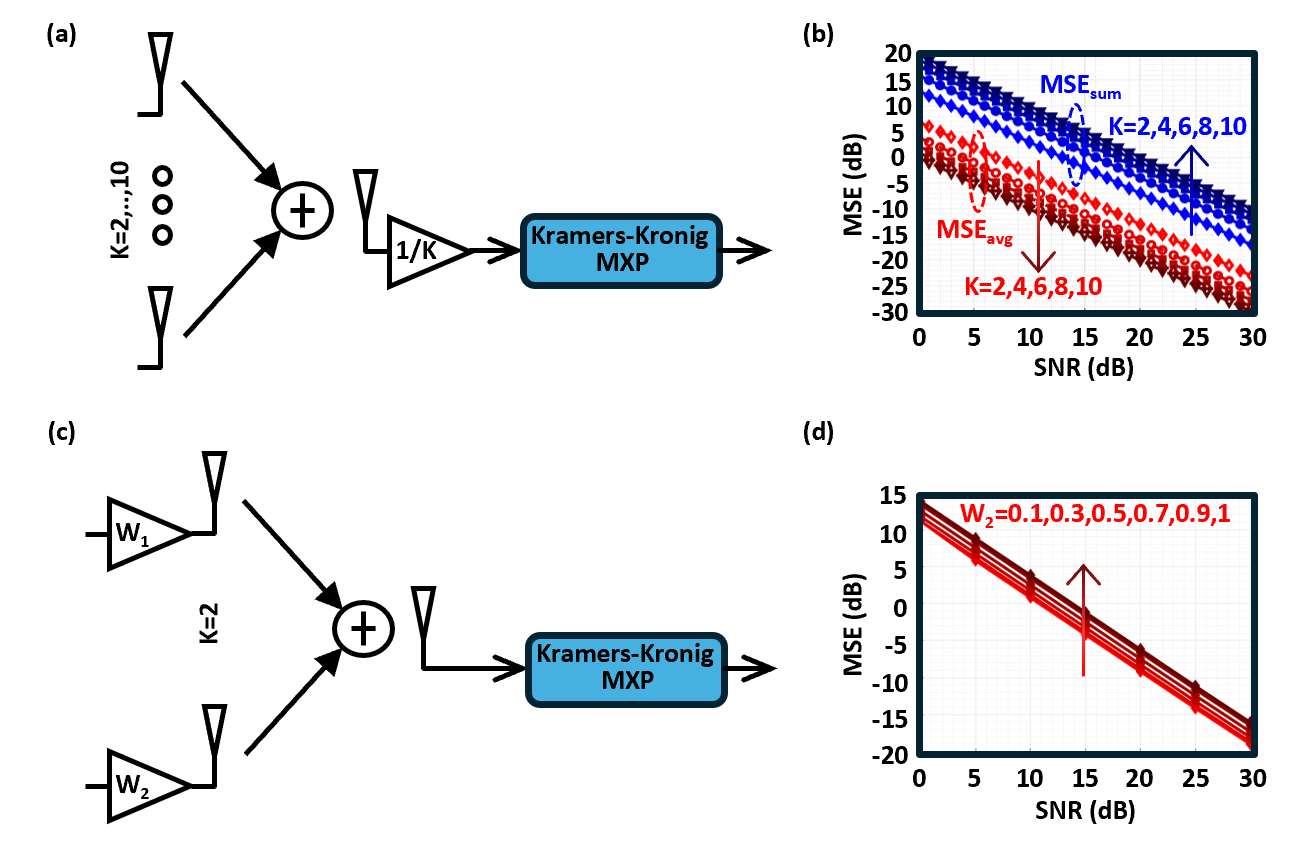}
 \caption{Over-the-air computation performance at 10~W/user. 
(a) Pre-/post-processing maps for sum/mean with $K\in\{2,\dots,10\}$. 
(b) MSE (dB) vs. SNR (0--30~dB) for $K=\{2,4,6,8,10\}$. 
(c) Maps for weighted sum (two users). 
(d) MSE (dB) vs. SNR with $W_1=1$ and $W_2=\{0.1,0.3,0.5,0.7,0.9,1\}$.}
  \label{MSEvsSNR}
\end{figure}

\subsection{Comparison between Coherent and Self-coherent Architectures}

Before introducing the formal system model, we compare conventional coherent and self-coherent KK-based OAC receivers, shown in Fig.~\ref{SystemModel}(a) and (b), respectively. In both cases, users transmit simultaneously over a multiple-access channel and their signals superpose at the receiver. Residual timing offsets are handled using matched-filter-based alignment at the receiver output, as in \cite{shao2021federated} (highlighted in blue in Fig.~\ref{SystemModel}), enabling symbol-level alignment prior to aggregation. Fig.~\ref{SystemModel} illustrates a two-user example with arbitrary 16-QAM symbols and the resulting summed constellation. Throughout this section, we assume perfect channel amplitude and phase estimation with pre-transmission channel inversion, such that the receiver observes a coherent sum. Key system parameters are summarized in Table~\ref{tab:sim_params}.

In the conventional direct-conversion architecture (Fig.~\ref{SystemModel}(a)), matched-filter alignment requires accurate carrier and timing synchronization between transmitters and receiver. While often assumed, this is difficult to achieve in practice, particularly at RF and mmWave frequencies where low-phase-noise PLLs and accurate quadrature generation are challenging. Carrier frequency offset (CFO) and phase offset (PO) between transmitter and receiver local oscillators directly distort the recovered baseband symbols, degrading coherent aggregation. These transmitter--receiver synchronization requirements, highlighted in red in Fig.~\ref{SystemModel}(a), constitute a fundamental bottleneck for scalable high-frequency OAC systems.

The self-coherent KK architecture (Fig.~\ref{SystemModel}(b)) relaxes this constraint by embedding a strong reference carrier in the transmitted spectrum. For sufficiently large carrier-to-signal power ratio (CSPR), the Kramers--Kronig mixed-signal processing (MXP) block reconstructs a faithful replica of the superposed signal at a lower intermediate frequency (IF). The reconstructed signal is self-synchronous, depending only on the mutual coherence between the transmitted reference carrier and RF signal frequencies $(F_{\mathrm{ref}}, F_{\mathrm{RF}})$, thereby eliminating transmitter--receiver CFO and phase offsets prior to baseband processing. Subsequent IF-to-baseband conversion can then be performed using a low-frequency PLL, simplifying quadrature generation and the matched-filter alignment stage.

The two architectures also differ in their mixing mechanisms. Conventional receivers perform frequency translation via multiplicative mixing with a local oscillator. In contrast, the KK MXP exploits nonlinear harmonic generation, particularly second-order terms, to enable down-conversion in the presence of the reference carrier. This process introduces self-mixing distortion intrinsic to KK reception, which is subsequently removed through nonlinear signal processing prior to final IF quadrature down-conversion.

\begin{table}[t]
\centering
\caption{Examples of nomographic functions for $K$ users. Each user $k$ applies a pre-processing map $\varphi_k(\cdot)$ to $s_k$, and the receiver computes
\(
f(s_1,\ldots,s_K)
=\psi\!\left(\sum_{k=1}^K \varphi_k(s_k)\right),
\)
i.e., aggregation of $\varphi_k(s_k)$ followed by post-processing $\psi(\cdot)$.}
\label{tab:nomographic_examples}
\vspace{0.5em}
\renewcommand{\arraystretch}{1.35}
\begin{tabular}{l|c|c|c}
\hline
\textbf{Case} & $f(s_1,\ldots,s_K)$ & $\varphi_k(s)$ & $\psi(s)$ \\
\hline
Arithmetic mean
& $\frac{1}{K}\sum_{k=1}^K s_k$
& $s$
& $\frac{s}{K}$ \\

Weighted sum
& $\sum_{k=1}^K w_k s_k$
& $w_k s$
& $s$ \\

\hline
\end{tabular}
\end{table}
\subsection{Nomographic Function Support and Aggregation Performance}

Having established the architectural differences between coherent and self-coherent receivers, we now examine the class of functions supported by the proposed KK-based framework and its achievable aggregation performance.

The self-coherent KK receiver supports a broad class of nomographic functions, summarized in Table~\ref{tab:nomographic_examples}. Each user $k$ applies a pre-processing map $\varphi_k(\cdot)$ to its local scalar $s_k$, and the receiver computes $f(s_1,\ldots,s_K)=\psi\left(\sum_{k=1}^K \varphi_k(s_k)\right)$, i.e., aggregation is performed over the air followed by a deterministic post-processing map $\psi(\cdot)$. This formulation includes key operations such as arithmetic mean and weighted sum, which are central to distributed learning, federated optimization, and collaborative sensing.

In the proposed architecture, KK-based mixed-signal processing reconstructs the superposed waveform prior to final baseband down-conversion. Consequently, these functions can be computed directly from the reconstructed intermediate-frequency (IF) signal without requiring coherent carrier-phase tracking between transmitters and receiver, removing the dependence on tight synchronization while preserving linear superposition.

The achievable performance is illustrated in Fig.~\ref{MSEvsSNR}. At an average transmit power of 10~W per user, both sum/mean aggregation for varying numbers of users (Fig.~\ref{MSEvsSNR}(a)--(b)) and weighted-sum aggregation in the two-user case (Fig.~\ref{MSEvsSNR}(c)--(d)) closely approach the theoretical mean-squared error (MSE) limits of ideal baseband transmission over a wide SNR range. The MSE scales linearly with SNR (in dB), indicating that KK reconstruction and harmonic-based down-conversion do not introduce additional penalties beyond those of the additive noise channel.

These results demonstrate that the self-coherent architecture preserves the information-theoretic efficiency of baseband OAC while eliminating transmitter--receiver CFO sensitivity, achieving near-optimal aggregation accuracy without coherent carrier-phase recovery.

\subsection{Proposed Transmission Scheme}

\noindent \textbf{Network Model and Objective.}
We consider a short-range uplink multiple-access channel (MAC) consisting of $K$
edge devices and a single access point (AP), where all nodes are equipped with a
single antenna.
The communication range is sufficiently small (e.g., indoor or line-of-sight
dominated), so that channel variations are primarily phase-based rather than
fading-based.
Each device $k\in[K]$ holds a binary data input
$\mathbf{b}_k \in \{0,1\}^{L}$ of length $L$, which is already prepared for
transmission.
Encoding, compression, and quantization are assumed fixed and are not modeled
further in this section.
The objective of the receiver is to reliably recover the aggregate transmitted
signal, which can subsequently be used to compute a desired function.

\noindent \textbf{Link Budget Considerations.}
To assess the practical feasibility of the proposed architecture in
sub-THz wireless systems, we briefly examine the achievable link budget.
Operating around 100~GHz enables the use of wide available bandwidths
while benefiting from highly directional antennas commonly employed in
sub-THz links. Assuming a lens-based antenna configuration with a total
antenna gain of 106~dBi and a transmit power of $-16$~dBm, consistent
with prior sub-THz demonstrations~\cite{harter2020generalized}, the
measured receiver sensitivity of $-47$~dBm supports wireless link
distances on the order of kilometers while maintaining approximately
17~dB SNR over a 2~GHz bandwidth. These estimates indicate that the
proposed receiver architecture can support practical high-SNR
sub-THz links when combined with directional antenna front-ends.

\vspace{0.5em}
\noindent \textbf{Symbol Mapping.}
Each device maps its $L$-bit input into a single complex-valued channel symbol
$s_k \in \mathbb{C}$ using a fixed modulation scheme of order $M$
(e.g., $M$-QAM), where each symbol conveys $\log_2 M$ bits.
Accordingly, we assume $L = \log_2 M$.
In addition, the transmitted symbols are normalized such that
$\mathbb{E}[|s_k|^2] \le 1$.
The symbols $\{s_k\}_{k=1}^K$ are assumed mutually independent across devices.

\vspace{0.5em}
\noindent \textbf{Transmit Power Scaling.}
To mitigate channel amplitude variations while respecting a per-device average
power constraint $P$, each transmitter applies a real-valued scaling coefficient
\begin{align}
\beta_k \triangleq \min\!\left\{ |h_k|^{-1},\, \sqrt{P} \right\}.
\label{eq:power_scaling}
\end{align}

\vspace{0.5em}
\noindent \textbf{Transmit Waveform.}
Let $T_s$ denote the symbol period and let $p(t)$ be a unit-energy pulse shape. Each device $k$ transmits a single symbol over one symbol interval.
The complex baseband waveform transmitted by user $k$ is
\begin{align}
x_k(t) \triangleq \beta_k s_k\, p(t).
\end{align}
Since $\mathbb{E}[|s_k|^2]\le 1$ and $\beta_k\le \sqrt{P}$, the average transmit
energy per symbol satisfies the power constraint.

\vspace{0.5em}
\noindent \textbf{KK-Compatible Signal Construction.}
To enable KK phase reconstruction using direct detection, the
transmitted signal includes a strong real-valued bias (carrier)
\begin{align}
u(t) \triangleq A + \sum_{k=1}^K x_k(t), \qquad A>0,
\label{eq:kk_u}
\end{align}
chosen sufficiently large to enforce a minimum-phase envelope over the symbol
interval.
The corresponding passband signal at carrier frequency $f_c$ is
\begin{align}
s(t) = \Re\!\left\{ u(t)\, e^{j2\pi f_c t} \right\}.
\end{align}

\vspace{0.5em}
\noindent \textbf{Short-Range Channel and Oscillator Model.}
Due to the short communication range and limited scattering, we do not adopt a
Rayleigh fading model.
Instead, the effective channel from device $k$ to the AP is modeled as a
deterministic amplitude $|h_k|$ capturing path loss and hardware gain, combined
with a time-varying phase induced by oscillator mismatch.
Each device employs an independent local oscillator with carrier frequency
$\tilde{f}_{c}$, which may differ from the receiver carrier frequency $f_c$.
The resulting carrier-frequency offset (CFO) is
\begin{align}
\Delta f \triangleq \tilde{f}_{c}-f_c.
\end{align}
In addition, oscillator imperfections introduce stochastic phase noise.
Over a single symbol interval, the combined effect of CFO and phase noise is
captured by an effective random phase rotation.

\vspace{0.5em}
\noindent \textbf{Direct Detection and KK Reconstruction.}
At the receiver, the passband signal is converted to a complex baseband envelope
$r(t)$ and processed by a square-law detector, producing the received intensity
\begin{align}
z(t) \triangleq |r(t)|^2,
\qquad
\rho(t) \triangleq |r(t)| = \sqrt{z(t)}.
\end{align}
Under the minimum-phase condition enforced by the bias $A$, the KK relation
allows reconstruction of the signal phase from the logarithm of the magnitude.
Specifically, the reconstructed phase is given by
\begin{align}
\widehat{\varphi}(t)
\triangleq
\mathcal{H}\!\left\{ \log \rho(t) \right\},
\end{align}
where $\mathcal{H}\{\cdot\}$ denotes the Hilbert transform.
The KK-recovered complex baseband envelope is then
\begin{align}
\widehat{r}(t)
\triangleq
\rho(t)\, e^{j\widehat{\varphi}(t)}.
\end{align}
\begin{figure*}[!ht]
  \centering
  \includegraphics[width=.9\linewidth]{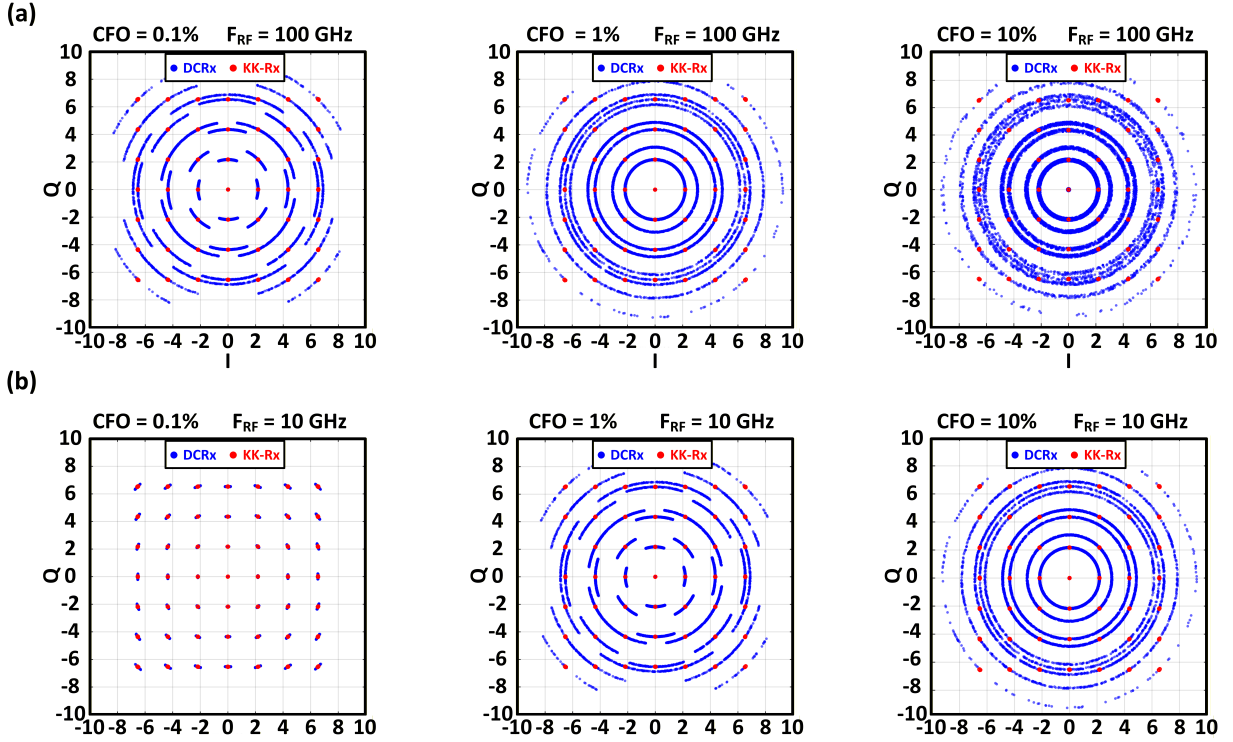}
  \caption {Impact of carrier frequency offset (CFO) between the transmitter and the receiver on the recovered constellation as a function of oscillator frequency accuracy and RF carrier frequency. A conventional direct-conversion receiver (blue) is compared against a self-coherent Kramers--Kronig receiver (red). (a) RF = 100 GHz. (b) RF = 10 GHz.}
  \label{ConstvsCFO}
\end{figure*}
\vspace{0.5em}
\noindent \textbf{Symbol-Rate Discrete-Time Equivalent Model.}
The recovered signal $\widehat{r}(t)$ is passed through a receive filter
(e.g., a matched filter with impulse response $p^{*}(-t)$) and sampled at the
symbol period $T_s$.
Since each device transmits a single symbol, the analysis focuses on one symbol
interval, and the resulting discrete-time observation is
\begin{align}
\widehat{y}
=
A\alpha_0
+
\sum_{k=1}^K |h_k| e^{j\phi_k}\, \beta_k s_k
+
\widetilde{n},
\label{eq:kk_discrete}
\end{align}
where $\phi_k$ denotes the effective residual phase associated with user $k$
after KK reconstruction (capturing the combined effect of CFO and phase noise),
$\alpha_0$ denotes the effective response of the bias term, and
$\widetilde{n}$ aggregates filtered additive white Gaussian noise, detector noise,
and residual KK reconstruction error.
Equation~\eqref{eq:kk_discrete} serves as the discrete-time equivalent model used
for subsequent analysis.

\section{Studying the Impact of CFO Impairment} 
Carrier-frequency offset (CFO) is a fundamental impairment in practical wireless systems arising from mismatches between transmitter and receiver oscillators. Its impact becomes increasingly severe at mmWave and sub-THz frequencies due to the scaling of oscillator inaccuracies with carrier frequency. In this section we analyze how CFO scales with RF frequency, examine its impact on conventional coherent receivers, and explain why the proposed self-coherent Kramers--Kronig architecture significantly relaxes these synchronization requirements.

\subsection{CFO Origin and Scaling with Carrier Frequency}
CFO in practical hardware is fundamentally dictated by oscillator frequency accuracy, which is commonly specified in parts-per-million (PPM). The achievable PPM depends strongly on oscillator type and implementation, most notably the quality of the crystal reference and whether temperature compensation is employed. In practice, state-of-the-art temperature-compensated crystal oscillators can achieve accuracies on the order of 0.1 PPM, whereas values around 5--10 PPM are more typical and generally considered acceptable for many RF systems.

For a given oscillator PPM, the resulting CFO expressed in absolute frequency units scales linearly with the RF carrier frequency, given by :
\begin{align}
\Delta f= F_{\text{RF}} \times \frac{\text{PPM}}{10^6}.
\label{eq:CFO}
\end{align}
It is worth highlighting that while operation at higher RF frequencies offers several compelling advantages---including access to larger contiguous bandwidths, improved integration and scaling to smaller form factors, and the ability to realize highly directional links using compact antenna arrays---it also exacerbates synchronization challenges. As the carrier frequency increases, even modest PPM values translate into large absolute CFOs, placing stringent requirements on carrier synchronization and directly impacting receiver robustness.

\subsection{Impact of CFO on Conventional Coherent Receivers}

In conventional direct-conversion receivers, the CFO between the transmitter and receiver local oscillators manifests as a residual frequency offset that directly degrades the recovered constellation through continuous phase rotation and inter-symbol interference. Since oscillator accuracy is typically specified in parts-per-million (PPM), the absolute CFO scales linearly with the RF carrier frequency according to (\ref{eq:CFO}). Consequently, the impact of CFO becomes increasingly severe as RF frequency increases.

This behavior is illustrated in Fig.~\ref{ConstvsCFO}, which compares the recovered constellations of a conventional direct-conversion receiver for different oscillator accuracies and RF carrier frequencies. At 100~GHz (Fig.~\ref{ConstvsCFO}(a)), even modest PPM values translate into large absolute CFOs, leading to substantial constellation distortion and degraded symbol recovery. In contrast, at 10~GHz (Fig.~\ref{ConstvsCFO}(b)), the smaller absolute CFO results in noticeably less degradation. The approximately tenfold increase in absolute CFO at 100~GHz highlights the growing sensitivity of conventional coherent receivers to oscillator inaccuracies as RF frequency scales.

\subsection{CFO Robustness of the Self-Coherent KK Receiver}

By contrast, the self-coherent Kramers--Kronig receiver largely eliminates the need for transmitter--receiver carrier synchronization. In this architecture, the dominant effective CFO is determined by the frequency offset between the transmitted RF carrier and the co-broadcast reference carrier, both generated within the transmitter. Synchronizing these two signals locally is significantly more tractable than synchronizing physically separate oscillators across a wireless link.

Moreover, the Kramers--Kronig reconstruction produces an output at a fixed and relatively low intermediate frequency determined by the separation between the reference and information-bearing carriers. The receiver-side IF local oscillator can therefore be readily synchronized to this fixed frequency, largely independent of the RF carrier frequency.

As reflected by the red constellations in Fig.~\ref{ConstvsCFO}, this self-coherent reconstruction makes the receiver effectively immune to transmitter--receiver CFO. Unlike the conventional architecture, the recovered constellation remains essentially unchanged across oscillator PPM values and RF carrier frequencies, demonstrating that accurate signal reconstruction can be maintained even at very high RF carriers.

\begin{figure*}[!ht]
  \centering
  \includegraphics[width=\linewidth]{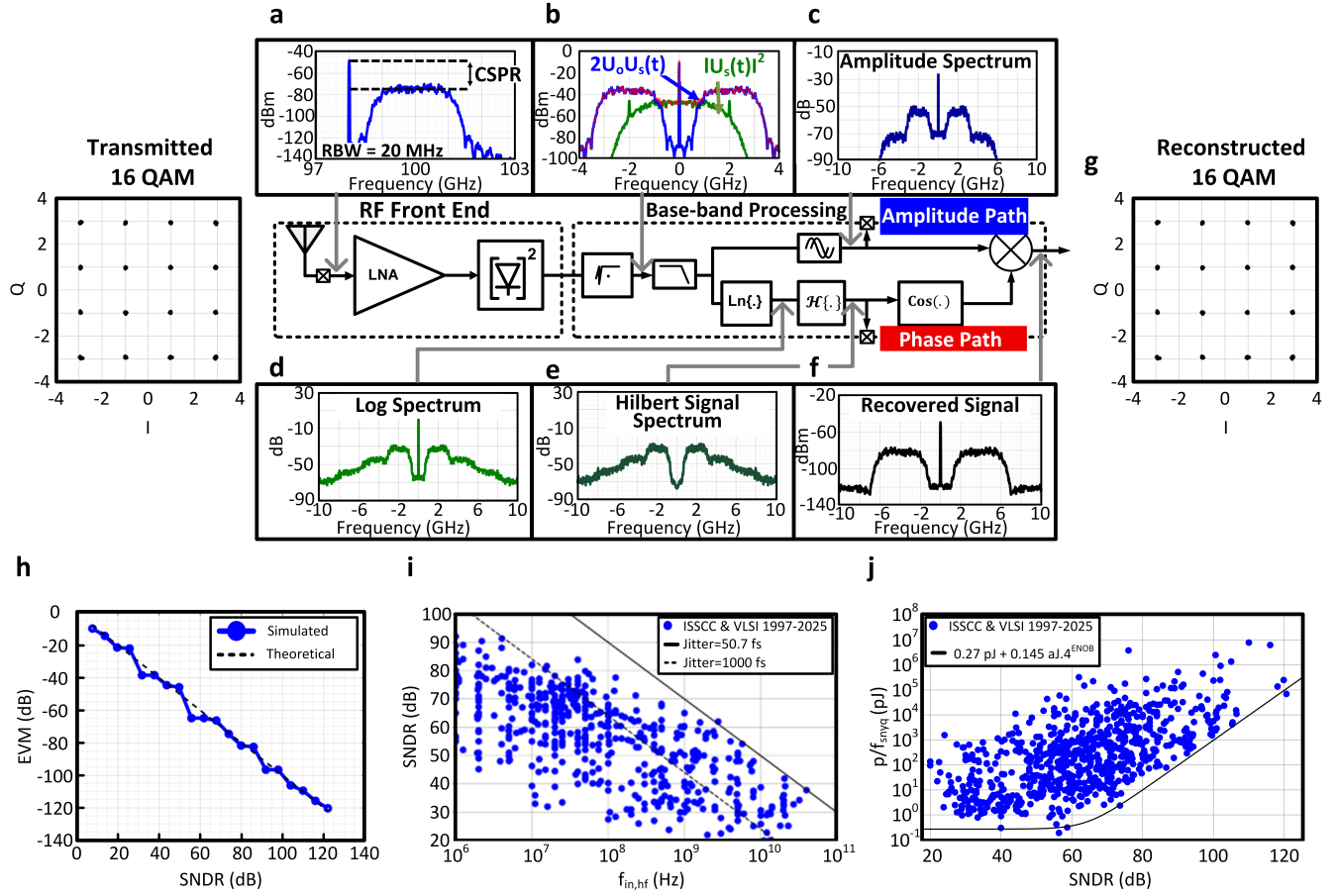}
  \caption{Signal processing in the Kramers--Kronig sub-THz receiver, showing amplitude (blue) and phase (red) paths. 
\textbf{(a)} Input spectrum 
\textbf{(b)} Rectifier output spectrum 
\textbf{(c)} Low-pass filtered output (recovered envelope) 
\textbf{(d)} Log-envelope spectrum 
\textbf{(e)} Hilbert transform of log-envelope 
\textbf{(f)} Reconstructed signal spectrum 
\textbf{(g)} Recovered constellation 
\textbf{(h)} RMS EVM vs.\ SNDR for quantized 16-QAM 
\textbf{(i)} ADC SNDR vs.\ Nyquist frequency ($f_{\text{in,NY}}$) from ISSCC \& VLSI (1997--2025), with jitter limits (50.7~fs, 1000~fs) 
\textbf{(j)} ADC energy efficiency (Walden FoM vs.\ SNDR) from ISSCC \& VLSI (1997--2025).}

\label{SystemSig}
\end{figure*}

\begin{figure}[!t]
\centering
\includegraphics[width=1\linewidth]{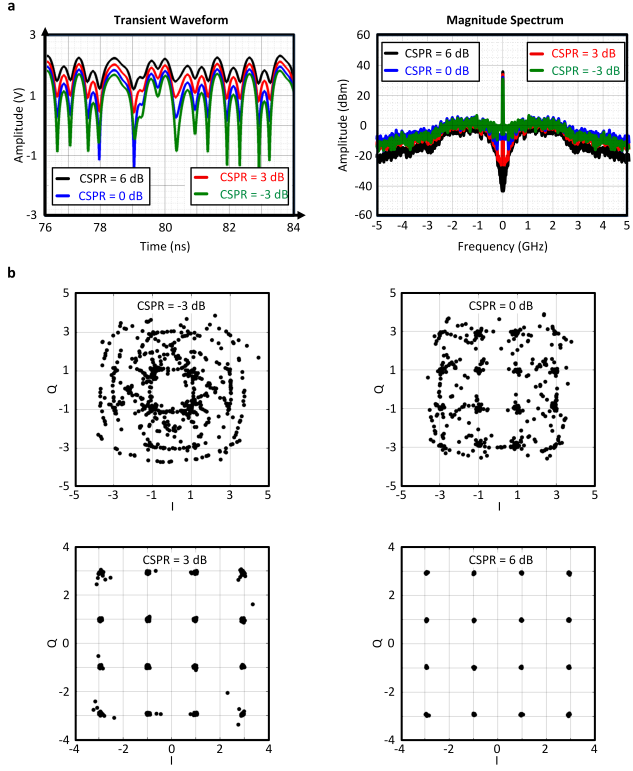}
\caption{Design trade-offs between error-vector-magnitude (EVM) and carrier-to-signal-power-ratio (CSPR). Effect of varying CSPR on \textbf{(a)} Time- and frequency-domain representations of the logarithmic envelope. \textbf{(b)} Recovered constellation.}
\label{ReconConst}
\end{figure}
\begin{figure}[!t]
\centering
\includegraphics[width=0.8\linewidth]{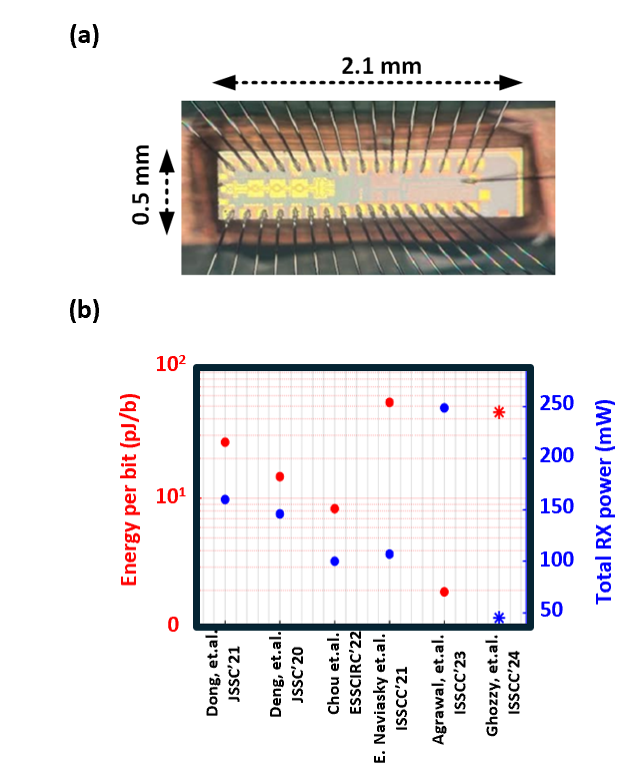}
\caption{(a) Micrograph of the 65-nm CMOS receiver (2.1\,mm $\times$ 0.5\,mm). 
(b) Energy efficiency (pJ/bit) and total power comparison with state-of-the-art sub-THz receivers, showing competitive efficiency at low power.}
\label{Eff_comp}
\end{figure}

\section{Receiver Reconstruction}
This section is organized into three parts. First, Section~\ref{sec:recon_analysis} examines the signal-reconstruction principle underlying the proposed receiver, including the parallel amplitude/phase processing paths, the minimum-phase and single-sideband conditions required for Kramers--Kronig recovery, and the resulting design trade-offs with carrier-to-signal power ratio. The following two sections then shift from analysis to implementation and experimental validation. Specifically, Section~\ref{sec:eneff} summarizes the fabricated chip and positions its power efficiency relative to prior state-of-the-art receivers, while Section~\ref{sec:measrslts} presents measured single-user results that validate broadband signal reconstruction and characterize sensitivity to input power and CSPR.
\subsection{Signal Path and Receiver Architecture}
\label{sec:recon_analysis}

The proposed integrated receiver employs a parallel signal-processing architecture for spectrally efficient sub-THz communication. The front end consists of a low-noise amplifier (LNA) whose output is split into two paths (Fig.~\ref{SystemSig}). The upper path performs envelope detection via rectification, followed by square-root and low-pass filtering to extract amplitude information. In parallel, the lower path computes the logarithm of the envelope and applies a Hilbert transform to recover phase. The two paths are then combined to reconstruct the modulated signal at an intermediate frequency (IF) without requiring an RF frequency synthesizer.

Accurate phase reconstruction requires that the incident RF waveform be single-sideband (SSB) and satisfy the minimum-phase (MP) condition, ensuring a unique phase--amplitude relationship~\cite{mecozzi2016kramers}. This is enforced by maintaining a sufficient carrier-to-signal power ratio (CSPR), preserving analyticity through the logarithmic operation and enabling phase recovery via Hilbert transformation.

The incident RF signal at the receiver input is
\begin{eqnarray}
U_{\mathrm{THz}}(t) = \Re\!\left\{U(t)e^{j2\pi f_{\mathrm{THz}}t}\right\},
\label{set4}
\end{eqnarray}
where the complex envelope is
\begin{eqnarray}
U(t) = U_0+U_s(t)=|U(t)|e^{j\phi(t)},
\label{set5}
\end{eqnarray}
with $\mathbf{U_o}$ denoting the carrier and $\mathbf{U_s(t)}$ the analytical SSB modulation. The MP condition is satisfied when~\cite{harter2020generalized}
\begin{eqnarray}
|U_0| \geq |U_s(t)|,
\label{MP}
\end{eqnarray}
ensuring the envelope never crosses zero. Taking the logarithm yields
\begin{eqnarray}
\ln U(t)=\ln|U(t)|+j\phi(t),
\label{u0}
\end{eqnarray}
so that phase is recovered via
\begin{eqnarray}
\phi(t)= \frac{1}{\pi}\mathcal{P} \int_{-\infty}^{\infty}\frac{\ln|U(\tau)|}{t-\tau}\,d\tau.
\label{phi}
\end{eqnarray}
Violating (\ref{MP}) causes envelope nulls, leading to logarithmic singularities, spectral spreading, and breakdown of the Hilbert relationship.

Fig.~\ref{SystemSig} illustrates signal evolution for a representative 16-QAM example occupying 99--101 GHz with $-54$ dBm power and 20 dB SNR. A $-48$ dBm carrier at 98 GHz yields a CSPR of 6 dB and total received power of $-47$ dBm. Since the LNA amplifies both components equally, the CSPR is preserved.

The rectifier output is
\begin{eqnarray}
|U(t)|^2 = |U_0|^2+2\Re\{U_0^* U_s(t)\}+|U_s(t)|^2,
\label{u2}
\end{eqnarray}
where $2\Re\{U_0^* U_s(t)\}$ produces the desired IF signal and $|U_s(t)|^2$ introduces self-mixing distortion. Applying a square-root yields
\begin{eqnarray}
|U(t)| = U_0\left[1+\frac{\Re\{U_s(t)\}}{U_0}+\frac{|U_s(t)|^2-\left(\Re\{U_s(t)\}\right)^2}{2U_0^2}+O(U_0^{-3})\right],
\label{set2}
\end{eqnarray}
where higher-order distortion terms decay with increasing carrier amplitude and are further suppressed by low-pass filtering.

Phase recovery is performed by applying the Hilbert transform to the logarithmic envelope. When the MP condition is satisfied (CSPR = 6 dB), the logarithmic spectrum remains SSB, enabling accurate phase reconstruction. The resulting IF signal spans 2 GHz bandwidth centered at 2 GHz, requiring wideband Hilbert filtering. The reconstructed spectrum eliminates self-mixing distortion, yielding a clean 16-QAM constellation with phase recovered entirely from amplitude.

Fully digital implementations based on envelope digitization and Hilbert processing are prohibitively power-intensive. Fig.~\ref{SystemSig}h shows that 16-QAM reconstruction requires 50--60 dB SNDR to avoid EVM degradation. As shown in Fig.~\ref{SystemSig}i~\cite{adc_survey}, such performance is achievable only at low GHz frequencies, while near tens of GHz, ADCs are limited to $\sim$36 dB SNDR. Moreover, Fig.~\ref{SystemSig}j~\cite{adc_survey} shows that energy per Nyquist sample increases sharply with bandwidth, reaching hundreds of milliwatts at multi-GHz rates and scaling to several watts at higher speeds~\cite{harter2020generalized}. These constraints make digital architectures impractical, motivating the proposed analog and mixed-signal approach.

Fig.~\ref{ReconConst} evaluates reconstruction versus CSPR. At low CSPR ($-3$ dB, 0 dB), the MP condition is violated, causing spectral spreading and ambiguous phase recovery. At 3 dB, reconstruction marginally satisfies the MP condition. At 6 dB, the logarithmic envelope fully preserves its SSB nature, enabling accurate phase recovery and a distortion-free constellation.

\subsection{Energy Efficiency}
\label{sec:eneff}

Figure~\ref{Eff_comp}(a) shows the micrograph of the fabricated receiver implemented in 65-nm CMOS, occupying a compact 2.1\,mm $\times$ 0.5\,mm footprint. The chip consumes 45.3\,mW from a 1.2\,V supply. The majority of the power is devoted to the RF front-end, while the remaining mixed-signal circuitry required for amplitude and phase reconstruction operates at relatively low power. This architecture enables energy-efficient reception without relying on conventional frequency synthesis or quadrature LO generation.

Figure~\ref{Eff_comp}(b) compares the energy efficiency and total receiver power consumption of the proposed self-coherent receiver against representative state-of-the-art wideband sub-THz receivers reported in prior work. Energy efficiency is evaluated in terms of energy per recovered information bit, while the total receiver power includes all RF, analog, and mixed-signal baseband circuits.

As shown in the figure, prior designs operating at multi-Gb/s data rates typically incur receiver power consumptions ranging from hundreds of milliwatts to over a watt, resulting in energy efficiencies spanning several tens to hundreds of picojoules per bit. In contrast, the proposed receiver achieves an energy efficiency of 45~pJ/bit at a 1~Gb/s data rate while consuming only 45.3\,mW of total receiver power. Despite eliminating conventional carrier recovery and frequency synthesis circuits, the proposed architecture achieves competitive energy efficiency relative to existing mmWave and sub-THz receiver implementations.

\begin{figure}[!t]
\centering
\includegraphics[width = 0.5\textwidth]{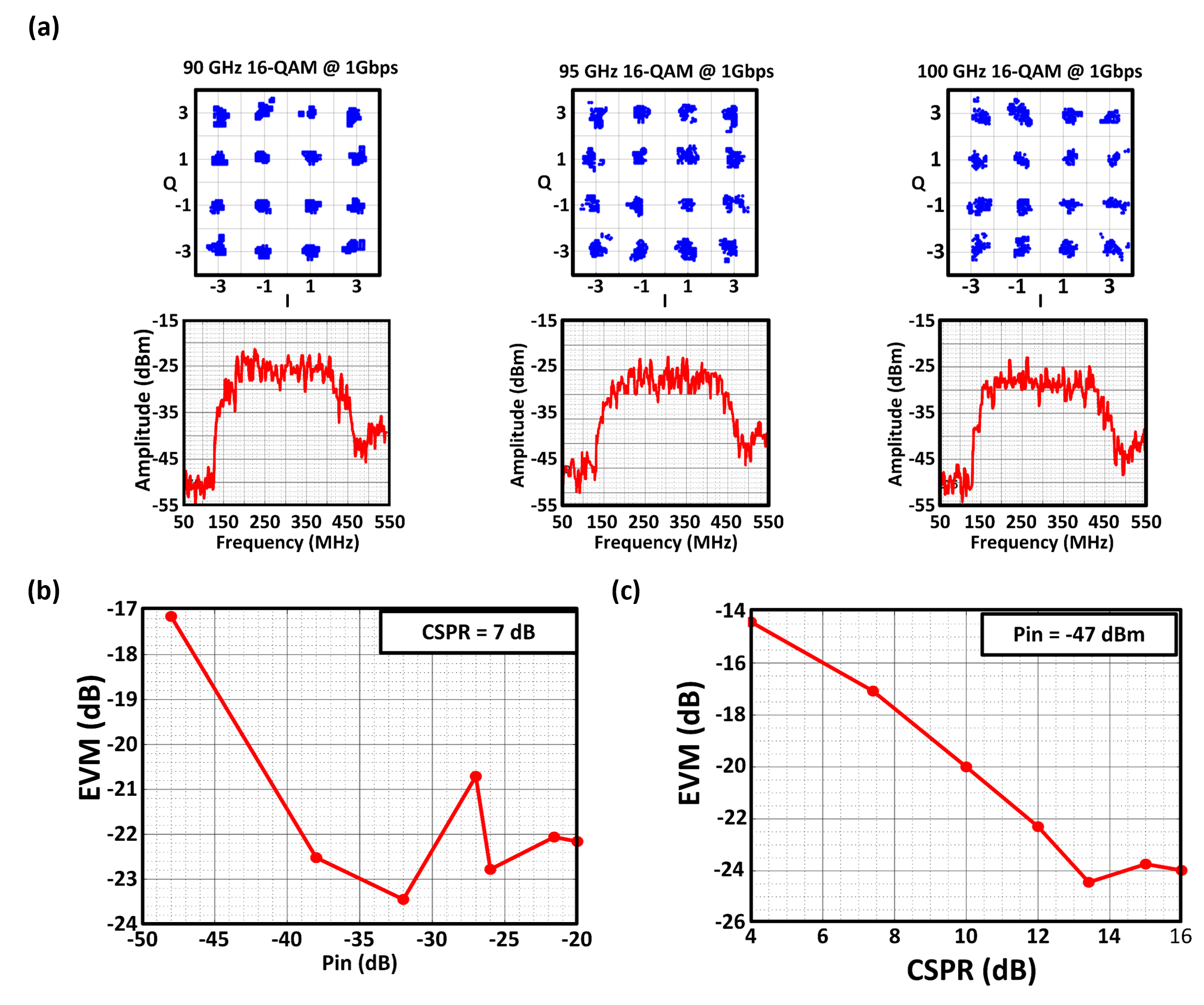}
\caption{Measured receiver performance. 
(a) Constellation and spectrum for 1~Gbps signals over 90--100~GHz, demonstrating successful reconstruction in a synthesizer-free architecture. 
(b) EVM vs. input power at $\mathrm{CSPR}=7~\mathrm{dB}$. 
(c) EVM vs. CSPR at $\mathrm{P}_{\mathrm{in}}=-47~\mathrm{dBm}$.}
\label{Sim_Meas_Proj2}
\end{figure}

\subsection{Single-User Measurement Results}
\label{sec:measrslts}

Figure~\ref{Sim_Meas_Proj2} summarizes the measured single-user performance of the receiver using 16-QAM modulation at 1\,Gb/s. Figure~\ref{Sim_Meas_Proj2}(a) shows reconstructed constellations and the corresponding output spectra for carrier frequencies of 90\,GHz, 95\,GHz, and 100\,GHz. The receiver directly reconstructs the signal at an intermediate frequency of approximately 300\,MHz with about 500\,MHz bandwidth. Across the measured carrier frequencies, the reconstructed constellations exhibit well-separated symbol clusters with error vector magnitudes (EVM) between approximately 5.4\% and 8\%, demonstrating stable broadband operation across the mmWave band without requiring frequency-dependent tuning.

The corresponding output spectra confirm correct recovery of the transmitted signal bandwidth. A mild spectral asymmetry is observed due to residual distortion introduced by the nonlinear envelope detection and reconstruction process. However, these distortion components are largely pushed outside the signal band by the analog processing chain and therefore do not significantly degrade the in-band signal quality required for reliable demodulation.

Figure~\ref{Sim_Meas_Proj2}(b) shows the measured EVM as a function of received input power while maintaining a fixed carrier-to-signal power ratio (CSPR) of 7\,dB. Reliable signal recovery is achieved down to an input power of approximately $-47$\,dBm, corresponding to an EVM of about $-17$\,dB. As the input power increases, the EVM improves and eventually saturates near $-22 \pm 1$\,dB for input power levels around $-25$ to $-20$\,dBm.

Figure~\ref{Sim_Meas_Proj2}(c) illustrates the dependence of reconstruction performance on CSPR at a fixed input power of $-47$\,dBm. At low CSPR values, the analyticity condition required by the Kramers--Kronig reconstruction is violated, leading to degraded phase recovery and increased EVM. Increasing the CSPR restores this condition and improves signal quality. The optimal operating region occurs around 13--15\,dB CSPR, where the measured EVM approaches $-24$\,dB.

Overall, these measurements demonstrate that gigabit-per-second mmWave links with higher-order modulation can be reliably recovered without explicit transmitter--receiver carrier synchronization or high-frequency local oscillators. By exploiting the analyticity condition through signal design and carrier placement, the proposed receiver enables low-power self-coherent reception, which is particularly attractive for scalable high-frequency wireless networks and distributed communication systems.

\section{Concluding Remarks}
\label{sec:conclusion}

This work introduced a self-coherent, synthesizer-free over-the-air computation framework based on Kramers--Kronig (KK) reception, addressing a key bottleneck in OAC systems: the need for accurate carrier synchronization and coherent reception. By combining biased aggregate transmission with direct detection and KK-based phase reconstruction, the proposed approach enables coherent-like aggregation with significantly reduced receiver complexity and synchronization overhead.

We developed a signal-domain model for multi-user OAC under KK reception and quantified aggregation distortion using a per-symbol MSE formulation. The analysis separates the effects of channel mismatch and KK reconstruction noise, and shows that higher-order envelope statistics govern the impact of modulation order.

Numerical results demonstrate robust aggregation performance without coherent synchronization. The dependence on modulation order is mild and can be beneficial due to averaging effects in function aggregation. These results position KK-based self-coherent reception as a promising solution for low-complexity, short-range, and high-frequency OAC systems. Future work includes extending the framework to frequency-selective channels, incorporating hardware impairments, and exploring applications in sub-THz and distributed learning scenarios.


\bibliographystyle{ACM-Reference-Format}
\bibliography{references}


\begin{thebibliography}{15}


\ifx \showCODEN    \undefined \def \showCODEN     #1{\unskip}     \fi
\ifx \showDOI      \undefined \def \showDOI       #1{#1}\fi
\ifx \showISBNx    \undefined \def \showISBNx     #1{\unskip}     \fi
\ifx \showISBNxiii \undefined \def \showISBNxiii  #1{\unskip}     \fi
\ifx \showISSN     \undefined \def \showISSN      #1{\unskip}     \fi
\ifx \showLCCN     \undefined \def \showLCCN      #1{\unskip}     \fi
\ifx \shownote     \undefined \def \shownote      #1{#1}          \fi
\ifx \showarticletitle \undefined \def \showarticletitle #1{#1}   \fi
\ifx \showURL      \undefined \def \showURL       {\relax}        \fi
\providecommand\bibfield[2]{#2}
\providecommand\bibinfo[2]{#2}
\providecommand\natexlab[1]{#1}
\providecommand\showeprint[2][]{arXiv:#2}

\bibitem[Dahl et~al\mbox{.}(2024)]%
        {dahl2024over2}
\bibfield{author}{\bibinfo{person}{Martin Dahl}, \bibinfo{person}{Zheng Chen}, {and} \bibinfo{person}{Erik~G Larsson}.} \bibinfo{year}{2024}\natexlab{}.
\newblock \showarticletitle{Over-the-air computation with reciprocity calibration: Detection and realignment of misaligned devices}. In \bibinfo{booktitle}{\emph{Proceedings of the 58th Asilomar Conference on Signals, Systems, and Computers.}} IEEE, \bibinfo{pages}{1832--1836}.
\newblock


\bibitem[Dahl and Larsson(2024)]%
        {dahl2024over}
\bibfield{author}{\bibinfo{person}{Martin Dahl} {and} \bibinfo{person}{Erik~G Larsson}.} \bibinfo{year}{2024}\natexlab{}.
\newblock \showarticletitle{Over-the-air federated learning with phase noise: Analysis and countermeasures}. In \bibinfo{booktitle}{\emph{Proceedings of the 58th Annual Conference on Information Sciences and Systems (CISS)}}. IEEE, \bibinfo{pages}{1--6}.
\newblock


\bibitem[Ghozzy et~al\mbox{.}(2024)]%
        {ghozzy202412}
\bibfield{author}{\bibinfo{person}{Sherif Ghozzy}, \bibinfo{person}{Muhamed Allam}, \bibinfo{person}{Emir~Ali Karahan}, \bibinfo{person}{Zheng Liu}, {and} \bibinfo{person}{Kaushik Sengupta}.} \bibinfo{year}{2024}\natexlab{}.
\newblock \showarticletitle{12.2 A mm-wave/sub-THz synthesizer-free coherent receiver with phase reconstruction through mixed-signal Kramer-Kronig processing}. In \bibinfo{booktitle}{\emph{Proceedings of the IEEE International Solid-State Circuits Conference (ISSCC)}}, Vol.~\bibinfo{volume}{67}. \bibinfo{pages}{220--222}.
\newblock


\bibitem[Harter et~al\mbox{.}(2020)]%
        {harter2020generalized}
\bibfield{author}{\bibinfo{person}{Tobias Harter}, \bibinfo{person}{Christoph F{\"u}llner}, \bibinfo{person}{Juned~N Kemal}, \bibinfo{person}{Sandeep Ummethala}, \bibinfo{person}{Johannes~L Steinmann}, \bibinfo{person}{Miriam Brosi}, \bibinfo{person}{Jeffrey~L Hesler}, \bibinfo{person}{Erik Br{\"u}ndermann}, \bibinfo{person}{A-S M{\"u}ller}, \bibinfo{person}{Wolfgang Freude}, {et~al\mbox{.}}} \bibinfo{year}{2020}\natexlab{}.
\newblock \showarticletitle{Generalized Kramers--Kronig receiver for coherent terahertz communications}.
\newblock \bibinfo{journal}{\emph{Nature Photonics}} \bibinfo{volume}{14}, \bibinfo{number}{10} (\bibinfo{year}{2020}), \bibinfo{pages}{601--606}.
\newblock


\bibitem[Mecozzi et~al\mbox{.}(2016)]%
        {mecozzi2016kramers}
\bibfield{author}{\bibinfo{person}{Antonio Mecozzi}, \bibinfo{person}{Cristian Antonelli}, {and} \bibinfo{person}{Mark Shtaif}.} \bibinfo{year}{2016}\natexlab{}.
\newblock \showarticletitle{Kramers--Kronig coherent receiver}.
\newblock \bibinfo{journal}{\emph{Optica}} \bibinfo{volume}{3}, \bibinfo{number}{11} (\bibinfo{year}{2016}), \bibinfo{pages}{1220--1227}.
\newblock


\bibitem[Murmann(2025)]%
        {adc_survey}
\bibfield{author}{\bibinfo{person}{Boris Murmann}.} \bibinfo{year}{2025}\natexlab{}.
\newblock \bibinfo{title}{{ADC Performance Survey 1997-2025}}.
\newblock
\newblock
\newblock
\shownote{[Online]. Available: https://github.com/bmurmann/ADC-survey}.


\bibitem[Pradhan et~al\mbox{.}(2025)]%
        {pradhan2025experimental}
\bibfield{author}{\bibinfo{person}{Suyash Pradhan}, \bibinfo{person}{Asil Koc}, \bibinfo{person}{Kubra Alemdar}, \bibinfo{person}{Mohamed~Amine Arfaoui}, \bibinfo{person}{Philip Pietraski}, \bibinfo{person}{Francois Periard}, \bibinfo{person}{Guodong Zhang}, \bibinfo{person}{Mario Hudon}, {and} \bibinfo{person}{Kaushik Chowdhury}.} \bibinfo{year}{2025}\natexlab{}.
\newblock \showarticletitle{Experimental demonstration of over the air federated learning for cellular networks}. In \bibinfo{booktitle}{\emph{2025 IEEE International Conference on Machine Learning for Communication and Networking (ICMLCN)}}. IEEE, \bibinfo{pages}{1--7}.
\newblock


\bibitem[{\c{S}}ahin(2025)]%
        {csahin2025feasibility}
\bibfield{author}{\bibinfo{person}{Alphan {\c{S}}ahin}.} \bibinfo{year}{2025}\natexlab{}.
\newblock \showarticletitle{On the feasibility of distributed phase synchronization for coherent signal superposition}. In \bibinfo{booktitle}{\emph{Proceedings of the International Symposium on Personal, Indoor and Mobile Radio Communications (PIMRC)}}. IEEE, \bibinfo{pages}{1--6}.
\newblock


\bibitem[{\c{S}}ahin et~al\mbox{.}(2021)]%
        {csahin2021distributed}
\bibfield{author}{\bibinfo{person}{Alphan {\c{S}}ahin}, \bibinfo{person}{Bryson Everette}, {and} \bibinfo{person}{Safi Shams~Muhtasimul Hoque}.} \bibinfo{year}{2021}\natexlab{}.
\newblock \showarticletitle{Distributed learning over a wireless network with FSK-based majority vote}. In \bibinfo{booktitle}{\emph{Proceeding of the 4th International Conference on Advanced Communication Technologies and Networking (CommNet)}}. \bibinfo{pages}{1--9}.
\newblock


\bibitem[{\c{S}}ahin and Yang(2023)]%
        {csahin2023survey}
\bibfield{author}{\bibinfo{person}{Alphan {\c{S}}ahin} {and} \bibinfo{person}{Rui Yang}.} \bibinfo{year}{2023}\natexlab{}.
\newblock \showarticletitle{A survey on over-the-air computation}.
\newblock \bibinfo{journal}{\emph{IEEE Communications Surveys \& Tutorials}} \bibinfo{volume}{25}, \bibinfo{number}{3} (\bibinfo{year}{2023}), \bibinfo{pages}{1877--1908}.
\newblock


\bibitem[Seif et~al\mbox{.}(2025)]%
        {seif2025collaborative}
\bibfield{author}{\bibinfo{person}{Mohamed Seif}, \bibinfo{person}{Yuqi Nie}, \bibinfo{person}{Andrea~J Goldsmith}, {and} \bibinfo{person}{H~Vincent Poor}.} \bibinfo{year}{2025}\natexlab{}.
\newblock \showarticletitle{Collaborative inference over wireless channels with feature differential privacy}.
\newblock \bibinfo{journal}{\emph{IEEE Journal on Selected Areas in Communications}} (\bibinfo{year}{2025}).
\newblock


\bibitem[Shao et~al\mbox{.}(2021)]%
        {shao2021federated}
\bibfield{author}{\bibinfo{person}{Yulin Shao}, \bibinfo{person}{Deniz G{\"u}nd{\"u}z}, {and} \bibinfo{person}{Soung~Chang Liew}.} \bibinfo{year}{2021}\natexlab{}.
\newblock \showarticletitle{Federated edge learning with misaligned over-the-air computation}.
\newblock \bibinfo{journal}{\emph{IEEE Transactions on Wireless Communications}} \bibinfo{volume}{21}, \bibinfo{number}{6} (\bibinfo{year}{2021}), \bibinfo{pages}{3951--3964}.
\newblock


\bibitem[Voelcker(2003)]%
        {voelcker2003demodulation}
\bibfield{author}{\bibinfo{person}{H Voelcker}.} \bibinfo{year}{2003}\natexlab{}.
\newblock \showarticletitle{Demodulation of single-sideband signals via envelope detection}.
\newblock \bibinfo{journal}{\emph{IEEE Transactions on Communication Technology}} \bibinfo{volume}{14}, \bibinfo{number}{1} (\bibinfo{year}{2003}), \bibinfo{pages}{22--30}.
\newblock


\bibitem[You et~al\mbox{.}(2023)]%
        {you2023broadband}
\bibfield{author}{\bibinfo{person}{Lizhao You}, \bibinfo{person}{Xinbo Zhao}, \bibinfo{person}{Rui Cao}, \bibinfo{person}{Yulin Shao}, {and} \bibinfo{person}{Liqun Fu}.} \bibinfo{year}{2023}\natexlab{}.
\newblock \showarticletitle{Broadband digital over-the-air computation for wireless federated edge learning}.
\newblock \bibinfo{journal}{\emph{IEEE Transactions on Mobile Computing}} \bibinfo{volume}{23}, \bibinfo{number}{5} (\bibinfo{year}{2023}), \bibinfo{pages}{5212--5228}.
\newblock


\bibitem[Zhu et~al\mbox{.}(2020)]%
        {zhu2020one}
\bibfield{author}{\bibinfo{person}{Guangxu Zhu}, \bibinfo{person}{Yuqing Du}, \bibinfo{person}{Deniz G{\"u}nd{\"u}z}, {and} \bibinfo{person}{Kaibin Huang}.} \bibinfo{year}{2020}\natexlab{}.
\newblock \showarticletitle{One-bit over-the-air aggregation for communication-efficient federated edge learning: Design and convergence analysis}.
\newblock \bibinfo{journal}{\emph{IEEE Transactions on Wireless Communications}} \bibinfo{volume}{20}, \bibinfo{number}{3} (\bibinfo{year}{2020}), \bibinfo{pages}{2120--2135}.
\newblock


\end{thebibliography}

\end{document}